\documentclass[12pt]{article} %***
\usepackage[sectionbib]{natbib}
\usepackage{array,epsfig,fancyheadings,rotating}
\usepackage[]{hyperref}  %<----modified by Ivan
\usepackage{sectsty, secdot,comment}
\sectionfont{\fontsize{12}{14pt plus.8pt minus .6pt}\selectfont}
\renewcommand{\theequation}{\thesection\arabic{equation}}
\subsectionfont{\fontsize{12}{14pt plus.8pt minus .6pt}\selectfont}

\usepackage{amsmath}
\usepackage{amssymb}
\usepackage{amsfonts}
\usepackage{multirow}
\usepackage{arydshln}
\usepackage{amsthm}
\usepackage{color}
\usepackage{threeparttable}

\newtheorem{theorem}{Theorem}
\newtheorem{lemma}{Lemma}
\newtheorem{coro}{Corollary}
\newtheorem{pro}{Proposition}
\theoremstyle{definition}
\newtheorem{defi}{Definition}
\newtheorem{example}{Example}
\newtheorem{remark}{Remark}
\newtheorem{construction}{Construction}
\newtheorem{case}{Case}
\newcommand{\ba}{\begin{array}}
\newcommand{\ea}{\end{array}}
\newcommand{\bt}{\begin{tabular}}
\newcommand{\et}{\end{tabular}}
\newcommand{\btb}{\begin{table}}
\newcommand{\etb}{\end{table}}
\newcommand{\bc}{\begin{center}}
\newcommand{\ec}{\end{center}}
\newcommand{\bea}{\begin{eqnarray}}
\newcommand{\eea}{\end{eqnarray}}
\newcommand{\Bea}{\begin{eqnarray*}}
\newcommand{\Eea}{\end{eqnarray*}}
\newcommand{\beq}{\begin{equation}}
\newcommand{\eeq}{\end{equation}}

\def \bfm#1{\mbox{\boldmath$#1$}}

\def \A {{\bfm A}} \def \a {{\bfm a}}  \def \b {{\bfm b}}
\def \B {{\bfm B}} \def \c {{\bfm c}} \def \C {{\bfm C}}
\def \d {{\bfm d}} \def \D {{\bfm D}} 
\def \E {{\bfm E}} 
 \def \G {{\bfm G}} \def \g {{\bfm g}}

  \def \0 {{\bfm 0}}
  
 \def \T {{\bfm T}}
\def \r {{\bfm r}} \def \R {{\bfm R}}
 
\def \t {{\bfm t}}  
\def \w {{\bfm w}}  \def \v {{\bfm v}}

\def \x {{\bfm x}} 
 \def \y {{\bfm y}}
 \def \z {{\bfm z}}

\def \one {{\bf 1}} \def \zero {{\bf 0}}
 
\def \Xi {{\bfm \xi}}

\begin{document}
%%%%%%%%%%%%%%%%%%%%%%%%%%%%%%%%%%%%%%%%%%%%%%%%%%%%%%%%%%%%%%%%%%%%%%%%%%%%%%%%%%%%%%%%%%%%%%%%%%%%%%%%%%%%%%%%%%%%%%%%%%%%
%%%%%%%%%%%%%%%%%%%%%%%%%%%%%%%%%%%%%%%%%%%%%%%%%%%%%%%%%%%%%%%%%%%%%%%%%%%%%%%%%%%%%%%%%%%%%%%%%%%%%%%%%%%%%%%%%%%%%%%%%%%%

\renewcommand{\baselinestretch}{2}

\markright{ \hbox{\footnotesize\rm Statistica Sinica
%{\footnotesize\bf 24} (201?), 000-000
}\hfill\\[-13pt]
\hbox{\footnotesize\rm
%\href{http://dx.doi.org/10.5705/ss.20??.???}{doi:http://dx.doi.org/10.5705/ss.20??.???}
}\hfill }

%\markboth{\hfill{\footnotesize\rm Feng Yang and C. Devon Lin} \hfill}
%{\hfill {\footnotesize\rm DOUBLY COUPLED DESIGNS FOR COMPUTERS EXPERIMENTS} \hfill}
\markboth{\hfill{\footnotesize\rm Feng Yang et al.} \hfill}
{\hfill {\footnotesize\rm  } \hfill}

\renewcommand{\thefootnote}{}
$\ $\par

%%%%%%%%%%%%%%%%%%%%%%%%%%%%%%%%%%%%%%%%%%%%%%%%%%%%%%%%%%%%%%%%%%%%%%%%%%%%%%%%%%%%%%%%%%%%%%%%%%%%%%%%%%%%%%%%%%%%%%%%%%%%

\fontsize{12}{14pt plus.8pt minus .6pt}\selectfont \vspace{0.8pc}
\centerline{\large\bf Doubly Coupled Designs for Computer Experiments}
 \vspace{2pt} \centerline{\large\bf with both Qualitative and Quantitative Factors}
\vspace{.4cm} \centerline{Feng Yang, C. Devon Lin, Yongdao Zhou and Yuanzhen He} \vspace{.4cm}
\centerline{\it Sichuan Normal University, Queen's University, Nankai University }
 \vspace{2pt}  \centerline{\it and Beijing Normal University}
\vspace{.55cm} \fontsize{9}{11.5pt plus.8pt minus .6pt}\selectfont

%%%%%%%%%%%%%%%%%%%%%%%%%%%%%%%%%%%%%%%%%%%%%%%%%%%%%%%%%%%%%%%%%%%%%%%%%%%%%%%%%%%%%%%%%%%%%%%%%%%%%%%%%%%%%%%%%%%%%%%%%%%%

%\title{ {\Large \bf Doubly Coupled Designs for Computer Experiments with
%both Qualitative and Quantitative Factors}}
%
%\author{  Feng Yang$^{a,c}$,\ \ C. Devon Lin$^b$,\ \ Yongdao Zhou$^c$\thanks{Corresponding author. Email: ydzhou@nankai.edu.cn }, Yuanzhen He$^d$ \vspace{2mm}\\
%{\small \it $^a$School of Mathematical Sciences, Sichuan Normal University, Chengdu 610068, China}\\
%{\small \it $^b$Department of Mathematics and Statistics, Queen's University, Kingston, ON, K7L 3N6, Canada}\\
%{\small \it $^c$School of Statistics and Data Science, LPMC \emph{\&} KLMDASR, Nankai University, Tianjin
%300071, China}\\
%{\small \it $^d$Department of Statistics, Beijing Normal University, Beijing 100875, China}
%\date{} }

%\thanks{}
%\date{}
%\maketitle
%\noindent

\begin{quotation}
\noindent {\it Abstract:} Computer experiments with both qualitative and
quantitative input variables
occur frequently in many scientific and engineering applications.
How to choose input settings for  such
experiments is an important issue for accurate statistical analysis,
uncertainty quantification and decision making. Sliced Latin
hypercube designs are the first systematic approach
to address this issue.  However, it
 comes with the increasing cost associated with an increasing
 large number of level combinations of the qualitative factors.
 For the reason of run size economy, marginally coupled
 designs were proposed in which the design for
the  quantitative factors is a sliced Latin hypercube design
  with respect to each qualitative factor. The drawback
  of such designs is that the corresponding data may not be
   able to capture the  effects between any two (and more) qualitative
   factors and  quantitative factors. To balance the run
   size and design efficiency, we propose a new type of
   designs,  doubly coupled designs, where the design
   points for the quantitative factors form a sliced Latin hypercube
   design with respect to the levels of any qualitative factor
   and with respect to the level combinations of  any two
   qualitative factors, respectively. The proposed designs
   have the better stratification property
between the qualitative and quantitative factors compared with  marginally coupled designs.
The existence of the
proposed designs is established.  Several construction methods are introduced, and the properties of the resulting designs are also studied.

\vspace{9pt} \noindent {\it Key words and phrases:} completely
resolvable orthogonal array,  sliced Latin hypercube,
stratification.
\par
\end{quotation}\par

\def\thefigure{\arabic{figure}}
\def\thetable{\arabic{table}}

\renewcommand{\theequation}{\thesection.\arabic{equation}}

\fontsize{12}{14pt plus.8pt minus .6pt}\selectfont

\setcounter{section}{1} %***
\setcounter{equation}{0} %-1
%\fancyhf{}
\lhead[\footnotesize\thepage\fancyplain{}\leftmark]{}\rhead[]{\fancyplain{}\rightmark\footnotesize\thepage}%Put this line in Page 2

\noindent {\bf 1. Introduction}

%%%%%%%%%%%%%%%%%%%%%%%%%%%%%%%%%%%%%
%\section{Introduction}
Computer experiments are one of the efficient ways to represent the
real world complex systems and have been increasingly used in the
physical, engineering and social sciences \citep{SWN03,FLS05}.
{For recent work on computer experiments, refer to
\cite{CSD18}, \cite{WSLL18}, \cite{XX18}, \cite{WXX18},
\cite{HYLW21}, and reference therein.} One prevailing way to select
input settings for computer experiments is to use Latin hypercube
designs (LHDs) proposed by \cite{MBC79}, because of the desirable
feature that when projected onto any factor, the resulting design
points spread out uniformly and achieve the maximum stratification.
{An LHD is not guaranteed to be space-filling in two or higher
dimensions and thus some improved LHDs are discussed, such as maxmin
LHDs  \citep{MM95,JH08,WXX18}, orthogonal array-based LHDs \citep{T93}, orthogonal LHDs
\citep{GE14, ST17, LLT20}, and strong orthogonal arrays-based LHDs
\citep{HT13,ZT19, ST20, WYL21}. However, such designs can be only
used when all the factors are continuous or quantitative.} In some
applications, the qualitative factors  are inevitable by nature, and
play a crucial role in the study of complex systems
\citep{RFWB06,LB06,JD07,QWW08,HJM09,HSNB09,ZQZ11,HLLY16}.
Consequently, it calls for the designs for computer experiments
involving both qualitative and quantitative factors.

A sliced Latin hypercube design (SLHD)
introduced by \cite{Q12} is an LHD with the property that it can be divided into several slices, each of which constitutes a
smaller LHD. It
maintains the maximum one-dimensional stratification for the whole
design as well as each slice. The first systematic approach to accommodate both qualitative
and quantitative factors in computer experiments is
to use an SLHD  for the quantitative factors and a (fractional) factorial design for the qualitative factors, and  each slice for the quantitative factors corresponds
to a level combination of the qualitative factors.
 It is evident that the run sizes of SLHDs grow rapidly as the number of the level combinations of the qualitative
factors increases. That is, an SLHD may be suitable for the situations that the number of the level combinations of the qualitative factors is relatively small, or the experiment is not expensive to run. Inspired by this, \cite{DHL15} proposed marginally coupled designs (MCDs), where
 the design points for the quantitative factors form an SLHD with
respect to any qualitative factor.  For the construction of MCDs, refer to \cite{DHL15}, \cite{HLS17,HLSL17}, \cite{HLS19} {and \cite{ZYL21}}.

MCDs select input settings that have the desirable {stratification
 between each qualitative factor and all quantitative factors.} However,
 some MCDs may have poor design properties between multiple
 qualitative factors and all quantitative factors. Intuitively, such design properties are important to study the interaction effects between multiple qualitative factors and quantitative factors, thereby
 possibly affecting  the accuracy of
an emulator for the underlying computer
simulator. Suppose that there are three
qualitative factors, the kind of raw materials
(say, M1, M2 and M3), the shape of raw materials (such as, thick,
 medium and thin), and the type of catalysts (C1, C2 and C3)
as well as other quantitative factors in an experiment. It is
sensible to adopt a design, where for each kind, each shape or
each catalyst, the associated design for the quantitative
factors has a desirable space-filling property, and it would
be more desirable
if for each level combination of any two qualitative factors, like
(M1, thick),
the corresponding design points for the quantitative factors enjoy the
appealing  space-filling property, which can help understand the effect between any two qualitative factors and the quantitative factors.
 In this paper, we focus on designs with the appealing
stratification properties between every two qualitative factors
and all quantitative factors,
along with all the features of MCDs.
We call such designs {\em doubly coupled designs} (DCDs).

Like in an MCD, a DCD uses an LHD for the quantitative factors. In addition, this LHD not only satisfies the constraint that for each level of any qualitative
 factor, the corresponding design points for the quantitative factors form an LHD,  but also
 the constraint that for each level combination of any two qualitative factors,
 the corresponding design points for the quantitative factors
 form an LHD. In other words,  for a DCD,  with respect to each qualitative factor, the design for the quantitative factors is an SLHD, and
  with respect to any two qualitative factors,
 the design for the quantitative factor is also an SLHD.
 The concept of DCDs sounds straightforward,
 however, the construction procedure of DCDs is not trivial and cannot be achieved
 by the simple extensions of the constructions for MCDs.

The rest of this paper is organized as follows. Section \ref{sec_defi}
presents the
notation and the definitions of the relevant designs. The
theoretical results of the existence for the proposed designs are discussed
 in Section \ref{sec_exis}. Section
\ref{sec_con} provides three constructions for DCDs.
The last section presents the conclusions and discussion.
All proofs are given in the online supplementary material.

%in that the design points of quantitative factors
%form a sliced Latin hypercube design with respect to the level combinations
%of any two factors of the qualitative factors in the doubly coupled design.

%%%%%%%%%%%%%%%%%%%%%%%%%%%%%%%%%%%%%
\setcounter{equation}{0}
\section{Notation and Definitions}\label{sec_defi}

An $n\times m$ matrix, of which the $j$-th column has $s_j$ levels
$\{0,1,\dots,s_j-1\}$, is an orthogonal array
of $n$ rows, $m$ factors and
strength $t$, if each of all possible level combinations occurs with the same
frequency in any of its $n\times t$ submatrix. Such an array is denoted by
OA$(n,m,s_1\cdots s_m,t)$. If some of $s_i$'s are equal, denote it by
OA$(n,m,s_1^{u_1}\cdots s_l^{u_l},t)$, where $\sum_i^lu_i=m$.
Furthermore, if all of $s_i$'s are identical, denote it by OA$(n,m,s,t)$.
An OA$(n,m,s,2)$  is called a completely resolvable orthogonal array, denoted
by CROA$(n,m,s,2)$,
if its rows can be divided into ${n}/{s}$ subarrays, such that each of
which is an OA$(s,m,s,1)$.

A  Latin hypercube of $n$ rows and $m$ factors, denoted by LH($n,m$), is an $n \times m$ matrix, each column of which is {a  permutation} of the $n$ equally-spaced levels, say $\{ 0, 1, \ldots, n-1\}$.  Given a Latin hypercube $L=(l_{ij})$, a random Latin hypercube design $D=(d_{ij})$ can be generated by $d_{ij} = (l_{ij}+u_{ij})/n $ where $u_{ij}$ is a random number from (0,1).  A  Latin hypercube design possesses the property that each of the $n$ equally-spaced intervals has exactly one design point.
A random Latin hypercube design may not be space-filling in two or higher dimensional projections.
Orthogonal array-based Latin
hypercubes introduced by \cite{O92} and \cite{T93}  resolve this issue  and guarantee the same grids stratification in low dimensional projections  as the original orthogonal array.  We review the construction method here. Assume an OA$(n,m,s,t)$ exists.
For each column of the orthogonal array, replace  the $n/s$ positions of level $i$  by
a random permutation
of $\{i(n/s), i(n/s)+1,\dots, (i+1)(n/s)-1\},$ for $i=0,1,\dots,s-1$. The resulting design
is an LH$(n,m)$.
Throughout this paper,
we call this method as the level-expansion method. Conversely, an array can be obtained by replacing $\{i{(n/s)}, i{(n/s)}+1,\dots, (i+1){(n/s)}-1\}$
 to the integer  $i,$ for $i=0,\dots,s-1$, and this is referred to as the
 level-collapsion method.

%Given an LH($n,m$), a Latin hypercube design (LHD) on $(0,1)^m$ can be obtained by jittering the %design points in each of the $n$ equally-spaced intervals. More specifically, let $h_{ij}$ be the %$(i,j)$th element of an  LH($n,m$),  $h^*_{ij}$ be the $(i,j)$th element of the corresponding LHD, %and let
%$h^*_{ij} = (h_{ij} + u_{ij})/n$, where $u_{ij}$'s are independent random numbers from ($0,1$).

Let $\D_1$ and $\D_2$ be the $n$-run
designs  for $q$ qualitative factors and $p$ quantitative factors, respectively, and denote  $\D=(\D_1,\D_2)$.
A design $\D=(\D_1,\D_2)$ is called a {\em marginally coupled design}  if $\D_2$ is a Latin hypercube  and the rows in $\D_2$ corresponding
to each level of each factor  in $\D_1$ form a Latin hypercube design.

\iffalse{If the points of the quantitative factors have the desirable  stratification property with respect to the qualitative factors, we say that the design $\D$ possesses the desirable stratification property between the quantitative and qualitative factors.}\fi
MCDs possess the appealing stratification
property between each qualitative and all
quantitative factors. We extend the concept of MCDs, and introduce a general notion,  {\em $\omega$-way coupled designs}, which have
the stronger stratification
property between the two types of factors.

\begin{defi} \label{defi1} An $n$-run design $\D=(\D_1,\D_2)$
with $q$ s-level qualitative factors and $p$
quantitative factors,
 is called an $\omega$-way coupled design, if it satisfies: (i) $\D_2$ is an LH$(n,p)$; and
(ii) the rows in $\D_2$
corresponding to each  level
combination of any $l$ factors in $\D_1$ form an LHD, for $l=1, \ldots, \omega$.
\end{defi}

Clearly, an $\omega$-way coupled  design is also an $l$-way coupled design for any $l<\omega$. Besides, a one-way coupled design is exactly an MCD. In this paper, we focus on a  two-way coupled design and call it a {\em doubly coupled design}. We denote such
a design by DCD$(n,s^q,p)$. We concentrate on the study of DCDs with $\D_1$ being an ${\rm OA}(n,q,s,2)$.

\begin{figure} [!h]
\setlength{\abovecaptionskip}{0pt}
\centering
\includegraphics[width=5.2in, clip]{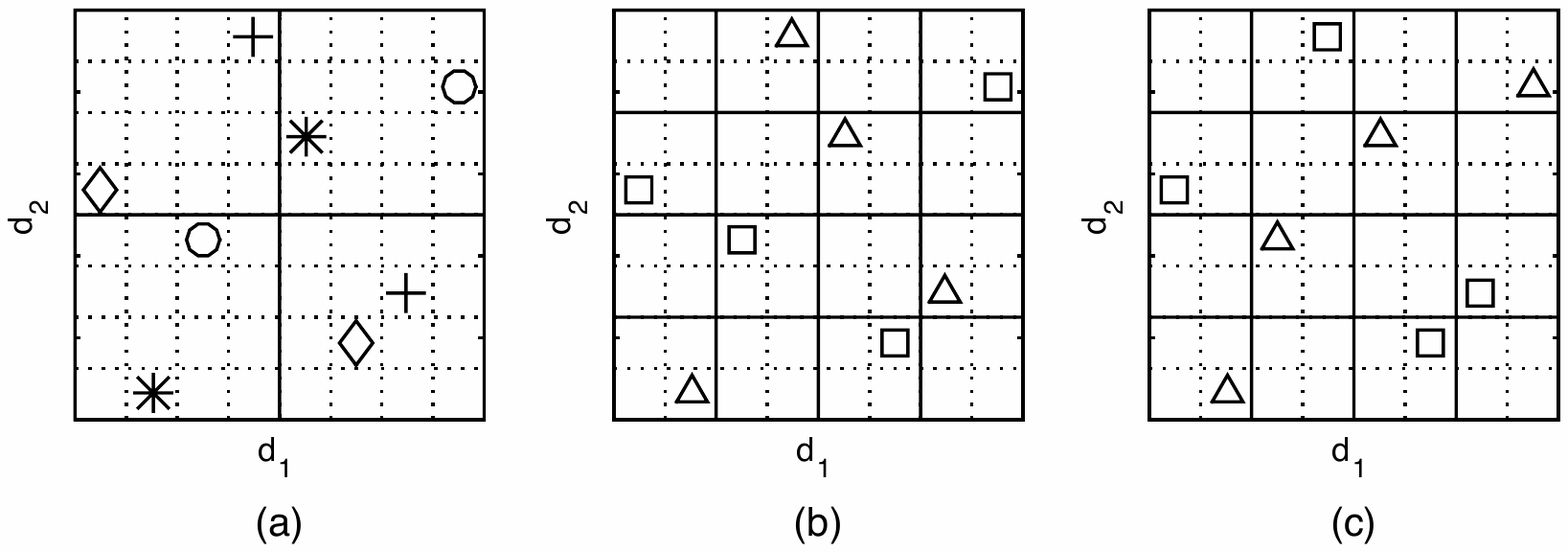}
\caption{Scatterplots of $\d_1$ versus $\d_2$ in Example \ref{ex1}:
(a) points represented by $\ast,+,\circ$ and $\lozenge$
correspond to the level combinations (0, 0), (0, 1), (1, 0), and (1, 1)
of factors $(\z_1,\z_2)$ ;
 (b) points marked by
 $\vartriangle$ and $\square$ correspond to the levels 0 and 1 of  $\z_1$;
 (c) points represented by $\vartriangle$ and $\square$
 correspond to the levels 0 and 1 of
$\z_2$.} \label{figure_ex1}
\end{figure}

Example \ref{ex1} below provides a DCD and its visualization.
\begin{example} \label{ex1} Consider the design $\D=(\D_1,\D_2)$ in Table \ref{table_ex1}. Let  $\z_1,\z_2$  be the two qualitative factors   and $\d_1,\d_2,\d_3,\d_4$ be the four quantitative factors.
%\begin{table}[!h] \caption{Design $\D=(\D_1,\D_2)$ in Example \ref{ex1}}\label{table_ex1} %!h means put the table here.
%\bc {\small \tabcolsep=8pt \bt {c|c} \hline $\D_1$ & $\D_2$
%\\\hline  $ \ba{cc}
%
%0& 0\\
%1& 1\\
%0& 1\\
%1& 0\\
%0& 0\\
%1& 1\\
%0& 1\\
%1& 0
% \ea  $ &  $\ba{cccc}
%1& 0&  0&  1\\
%0& 4&  4&  0\\
%6& 2&  6&  2\\
%7& 6&  2&  3\\
%4& 5&  5&  4\\
%5& 1&  1&  5\\
%3& 7&  3&  6\\
%2& 3&  7&  7
% \ea $  \\
% \hline
% \et} \ec
% \end{table}
 \begin{table}[!h]\caption{Design $\D=(\D_1,\D_2)$ in Example \ref{ex1}}\label{table_ex1}
\bc {\small  \bt {r|rrrrrrrr} \hline
\multirow{2}*{$\D_1^T$} & 0 & 1 & 0 & 1 & 0 & 1 & 0 & 1\\
                       ~& 0 & 1 & 1 & 0 & 0 & 1 & 1 & 0\\\hline
\multirow{4}*{$\D_2^T$} & 1& 0 & 6 & 7 & 4 & 5 & 3 & 2\\
~& 0 & 4 & 2 & 6 & 5 & 1 & 7 & 3\\
~& 0 & 4 & 6 & 2 & 5 & 1 & 3 & 7\\
~& 1 & 0 & 2 & 3 & 4 & 5 & 6 & 7\\\hline
 \et} \ec
\end{table}

Figures \ref{figure_ex1}(a), (b) and (c) display the design points for the first two quantitative factors $\d_1$ versus $\d_2$ with respect to the level combinations of $(\z_1,\z_2)$, the levels of $\z_1$ and the levels of $\z_2$, respectively. From Figure \ref{figure_ex1}(a), it is
apparent   that
the whole 8 points form an LHD while the points of $\D_2$ corresponding to
each of the four level combinations of ($\z_1,\z_2$)
 are LHDs with 2 levels, respectively. Figures \ref{figure_ex1}(b) and \ref{figure_ex1}(c) reveal that
the points in $\D_2$
corresponding to each level of $\z_1$ or $\z_2$  form an LHD, respectively. The plots for other quantitative dimensions are similar, so
 we omit them to save space.
 From Definition \ref{defi1}, it is a DCD$(8,2^2,4)$. Clearly,
 this DCD has the better stratification property between the qualitative and quantitative factors than an MCD,
since the design points for the quantitative factors in
 an MCD may not enjoy the maximum one-dimensional projection uniformity with respect to each level combination of any two
qualitative factors as Figure \ref{figure_ex1}(a).
\end{example}

%In fact, the operations in (\ref{collap1}) and
%(\ref{collap2}) are a application of level-collapsion method.
%and $\widetilde{\widetilde{D}}_2$ via level-expansion method.

%%%%%%%%%%%%%%%%%%%%%%%%%%%%%%%%%%%%%
\setcounter{equation}{0}
\section{Existence of DCDs}\label{sec_exis}

This section focuses on investigating the properties of DCDs and  establishing the existence of a DCD$(n,s^q,p)$, %in terms of the design properties of $\D_1$ and $\D_2$. The results are summarized in   Theorems \ref{iff2} and \ref{iff1},
which quantifies all the characteristics
of the sub-designs $\D_1$ and $\D_2$ in a DCD.

For ease of expression, more notations are introduced.
For
$\D_1=(\z_1,\dots,\z_q)$
and
$\D_2=(\d_1,\dots,\d_p)$ in a DCD, we
define
$\widetilde{\D}_2$ and $\widetilde{\widetilde{\D}}_2$ as,
 \bea \label{level_colla}
  \widetilde{\D}_2=\left\lfloor \frac{\D_2}{s}
 \right\rfloor=(\tilde{\d_1},\dots,\tilde{\d_{p}})
 ~\text{and} ~
 \widetilde{\widetilde{\D}}_2=\left\lfloor \frac{\widetilde{\D}_2}{s}
\right\rfloor=(\tilde{\tilde{\d_1}},\dots,\tilde{\tilde{\d_p}}),\eea where $\lfloor a\rfloor$
represents the largest integer not exceeding $a$. Since
$\D_2$ is an LH$(n,p)$, we have that
$\widetilde{\D}_2$ is an OA$(n,p,{n}/{s},1)$
and
$\widetilde{\widetilde{\D}}_2$ is an OA$(n,p,{n}/{s^2},1)$.
Conversely, $\D_2$  can be obtained from
$\widetilde{\D}_2$ via the
level-expansion method.

 Theorem \ref{iff2} provides the necessary and sufficient conditions on both
$\D_1$ and  $\D_2$ to ensure that a DCD exists.

\begin{theorem}\label{iff2}
Suppose  $\D_1=(\z_1,\dots,\z_q)$ is an OA$(n,q,s,2)$, and $\D_2=(\d_1,\dots,\d_p)$ is an LH$(n,p)$.
The design $\D=(\D_1,\D_2)$ is a DCD$(n,s^q,p)$ if and only if
\begin{enumerate}
\item[(a)] $(\z_i,\tilde{\d_k})$ is an OA$(n,2,s\left({n}/{s}\right),2)$,
for any $1 \leq i\leq q, 1\leq k\leq p$; and
\item[(b)] $(\z_i,\z_j,\tilde{\tilde{\d_k}})$ is an OA$(n,3,s^2\left({n}/{s^2}\right),3)$,  for any $1 \leq i\neq j\leq q, 1\leq k\leq p$.
\end{enumerate}

\end{theorem}

Condition (a) of Theorem \ref{iff2} is the necessary and
sufficient condition for $(\D_1,\D_2)$ to be an MCD, see \cite{HLS17}.
Condition  (b) says that for an MCD to be a DCD, $(\z_i,\z_j,\tilde{\tilde{\d_k}})$  must be a full factorial design.

In addition, it is worth noting that Conditions (a) and (b)
 are independent, that is, if a design satisfies Condition (a),
it may not meet Condition (b), vice versa. We give two
designs to illustrate this point. Let $\D^{(a)}
=(\D_1,\D_2^{(a)})$ and $\D^{(b)}=(\D_1,\D_2^{(b)})$,
where $\D_1$ is from Table \ref{table_ex1}, $
\D_2^{(a)}=((1,0,6,7,3,2,4,5)^T,(0,4,2,6,5,1,7,3)^T
)$ and $\D_2^{(b)}=\left((6,0,1,4,3,5,7,2)^T,\right.$ $\left.
(2,4,0,5,7,1,6,3)^T\right)$. It can be easily seen that
$\D^{(a)}$ meets Condition (a) but not (b), while $\D^{(b)}$
satisfies Condition (b) but not (a).

\begin{remark}\label{re1}  In Theorem \ref{iff2}, Condition (a)  indicates  that $(\D_1,\tilde{\d_k})$
is an OA$(n,q+1,s^{q}\left({n}/{s}\right),2)$. In addition, Condition (b) implies that $(\D_1, \tilde{\tilde{\d_k}})$
 is an OA$(n,q+1,s^q({n}/{s^2}),2)$.
\end{remark}

We now revisit Example \ref{ex1}  to show the
application of Theorem \ref{iff2}.
\begin{example}\label{ex_th1} (Example \ref{ex1}
continued) For the given $\D_2$, we can obtain $ \widetilde{\D}_2$ and
$\widetilde{\widetilde{\D}}_2$ via    (\ref{level_colla}) and we display these two designs as well as $\D_1$ in   Table \ref{table_ex_th1}.
 It can be checked that  $(\z_i,\tilde{\d_k})$ is an OA$(8,2,2\times4,2)$ and $(\z_i,\z_j,\tilde{\tilde{\d_k}})$ is an
OA$(8,3,2,3)$, for any $1\leq i\neq j\leq2$ and $1\leq k\leq4$.
 According to Theorem \ref{iff2}, the
design $\D$ in Example \ref{ex1} should be a
DCD$(8,2^2,4)$.

\begin{table}[!h] \caption{The $\D_1$, $ \widetilde{\D}_2$ and $\widetilde{\widetilde{\D}}_2$ in Example \ref{ex_th1}}\label{table_ex_th1} %!h means put the table here.
\bc {\small \tabcolsep=8pt \bt {c|c|c} \hline $\D_1$ & $\widetilde{\D}_2$ & $\widetilde{\widetilde{\D}}_2$
\\\hline  $ \ba{cc}

0&  0\\
1&  1\\
0&  1\\
1&  0\\
0&  0\\
1&  1\\
0&  1\\
1&  0
 \ea  $ &  $\ba{cccc}
0&  0&  0&  0\\
0&  2&  2&  0\\
3&  1&  3&  1\\
3&  3&  1&  1\\
2&  2&  2&  2\\
2&  0&  0&  2\\
1&  3&  1&  3\\
1&  1&  3&  3
 \ea $&  $\ba{cccc}
0&  0&  0&  0\\
0&  1&  1&  0\\
1&  0&  1&  0\\
1&  1&  0&  0\\
1&  1&  1&  1\\
1&  0&  0&  1\\
0&  1&  0&  1\\
0&  0&  1&  1
 \ea $  \\
 \hline
 \et} \ec
 \end{table}
\end{example}
Theorem \ref{iff2} establishes the existence of DCDs in terms of the relationship between the individual columns in $\D_1$ and $\tilde{\d_k}$, and the relationship between any pair of columns in $\D_1$ and $\tilde{\tilde{\d_k}}$. Interestingly, we can also give the existence of DCDs in terms of the design property of the entire design $\D_1$, which shows the required  structure of $\D_1$ in a DCD.
 The precise result is presented in Theorem \ref{iff1}.

\begin{theorem}\label{iff1}
A DCD$(n,s^q,p)$ exists if and only if
$\D_1$ can be partitioned into ${n}/{s^2}$  CROA$(s^2,q, s,2)$'s.
\end{theorem}

Theorem \ref{iff1} presents the requirement on $\D_1$ in
a DCD. In the construction of a
DCD, the $\D_1$ required by Theorem \ref{iff1} is the
cornerstone. Since for the given design parameters, only when
the
expected $\D_1$ exists we can construct the
corresponding $\D_2$  such that $\D=(\D_1,\D_2)$ is a DCD.

As an example of Theorem \ref{iff1}, see $\D_1$ in Table \ref{table_ex1}. The first four rows and the last four rows of $\D_1$ are CROA$(4,2,2,2)$'s, respectively.
The sufficiency of the proof in fact provides a procedure to construct $\D_2$'s.
The detailed process will be shown in Construction
\ref{con_dcd1} of Section \ref{sec_con}.

Theorem~\ref{ns_con} below studies the existence of a DCD in terms of the relationship between the columns of $\D_1$ and the columns of two relevant arrays that we use $\B$ and $\C$ to denote.

\begin{theorem}\label{ns_con}
Suppose  $\D_1$ is an OA$(n,q,s,2)$ and $\D_2$ is an LH$(n,p)$.
The design $\D=(\D_1,\D_2)$ is a DCD$(n,s^q,p)$ if and only if there exist two arrays, {$\B=OA(n, p, n/s^2, 1)$} and $\C=OA(n, p, s,1)$  such that  for  any $1\leq i\neq j\leq q$ and $1\leq k\leq
p$, both $({\z}_i, {\z}_j, {\b}_k)$ and $({\z}_i, {\c}_k, {\b}_k)$
 are OA$(n, 3, s^2\left({n}/{s^2}\right), 3)$'s, where $\z_i$ is the $i$th column of $\D_1$,  $\b_k$ and $\c_k$ are the $k$th column of $\B$ and $\C$, respectively,  and  $\widetilde {\D}_2$ in (\ref{level_colla}) can be written as
$\widetilde {\D}_2=s\B+\C$.
\end{theorem}

\begin{remark}\label{re_space}   The condition $\widetilde {\D}_2=s\B+\C$  in Theorem~\ref{ns_con}  implies $\widetilde{\widetilde{\D}}_2=\B$ which further implies that
the space-filling property of $\D_2$ heavily
relies on that of $\B$ and is slightly affected by $\C$. If $\B$
has a better stratification property, so is $\D_2$. For example,
 if $\B$ is an
 OA$(n,p,{n}/{s^2},2)$ instead of
OA$(n,p,{n}/{s^2},1)$,  $\D_2$ achieves the stratifications on $({n}/{s^2})\times ({n}/{s^2})$ grids for any two quantitative factors.

\end{remark}

Theorems \ref{iff2}, \ref{iff1} and \ref{ns_con} all  provide  necessary and sufficient conditions for a DCD to exist. These conditions are essentially the same but described in different ways for different purpose and usages. Theorem \ref{ns_con}  reveals that to construct a DCD, we shall find the $\D_1$, $\B$ and $\C$ that satisfy the conditions. Next section provides three ways to provide such $\D_1$, $\B$ and $\C$.

Before we move to the construction of DCDs in next section, we consider a theoretically and practically important topic in the study of DCDs, that is,  the maximum number
of $s$-level qualitative factors that an $n$-run DCD   can entertain. The following corollary gives the upper bound of the
qualitative factors in a DCD.

%This number also helps determine the minimum run size in a DCD to accommodate a given number of $s$-level qualitative factors.

\begin{coro}\label{coro1} If a DCD with $\D_1$ being an OA$(n,q,s,2)$ exists, then $q\leqslant s$.
\end{coro}

This proof of Corollary \ref{coro1} is straightforward by  Theorem \ref{iff1} and Lemma 1 of \cite{DHL15} and thus omitted. This corollary shows that the number of the qualitative factors in a DCD cannot exceed $s$.
Although the result seems restrictive, it is still practical. There are applications in the literature that the number of the qualitative
factors is no more than the  number of the qualitative levels,  for example, \cite{P89}
considered a router bit experiment with two qualitative four-level factors and seven quantitative factors. Moreover,
 when $s$ is a prime power, there always exists a CROA$(s^2,s,s,2)$ by deleting one column from the saturated OA$(s^2,s+1,s,2)$.  Stacking $n/s^2$ such CROAs to obtain a desired $\D_1$ in Theorem \ref{iff1}, the number of the qualitative factors of the resulting design $\D_1$ reaches the upper bound, $s$.

%Corollary \ref{coro1}  states that
%the number of qualitative factors accommodated in
%a DCD is no more than their levels.
%In most of practical experiments, the number of qualitative
%factors is relatively small and usually far less than that
%of quantitative factors, hence, the DCDs basically meet the
%actual demands. In addition,
%the examples in this paper
%demonstrate that this upper bound is tight since it can
% be reached readily.

%%%%%%%%%%%%%%%%%%%%%%%%%%%
 \section{Construction of DCDs}\label{sec_con}

For constructing DCDs, the computational search approach is often infeasible.
This section  presents three constructions to generate various DCDs.
The two methods in Subsection 4.1 %\ref{sub_con1}
construct DCDs by using the permutation approach, which can entertain a large number of the quantitative factors.
Subsection 4.2 %\ref{sub_con2}
can provide DCDs with the guaranteed projection space-filling properties on the quantitative factors,
while the number of the quantitative factors in DCDs may be relatively limited.
The constructions use orthogonal arrays of strength two or three which
are readily available in the textbooks such as \cite{HSS99} and the design catalogues on the websites such as \cite{S14}.

%\subsection{Constructions via the permutation method}\label{sub_con1}

\vspace{2mm}\noindent {\bf 4.1 Constructions of design $\D_2$ via permutations}\vspace{2mm}

In this subsection, we give two procedures based on permutations to construct DCDs with  a large number of  quantitative factors.

Let  $\A_1,\dots,\A_\lambda$ be OA$(s^2,q+1,s,2)$'s.
Without loss of generality, assume the last  column of every $\A_i$ is $(\zero_{s}^T, \one_{s}^T,\dots, {\bf (s-1)}_{s}^T)^T$, where $\y_{s}$ represents a column vector of length $s$ with all the element being $y$'s.  The $\lambda$ OAs will be used to generate  $\D_1$ in this subsection.

One construction procedure of $\D_1$, $\B$ and $\C$ in Theorem \ref{ns_con} works as follows and it uses the idea of the proof of Theorem \ref{iff1}.
 \begin{construction} \label{con_dcd1}
 ~
 \begin{itemize}
   \item[Step 1.] Obtain the array $\D_1$ by deleting the last column of $(\A_{1}^T, \ldots, \A_{\lambda}^T)^T$.
   \item[Step 2.]  Let $\b_k=\v_k \otimes  \one_{s^2}$, where $\v_k$ is a random permutation of $(0, 1, \dots, \lambda-1)^T$, for $1\leq k \leq p$,  and denote $\B=(\b_1,\dots,\b_p)$.
  \item[Step 3.]   Let ${\c}_k=(({\w}_{k1} \otimes {\bf 1}_s)^T, \ldots, ({\w}_{k\lambda}\otimes {\bf 1}_s)^T)^T$, where ${\w}_{kj}$ is a random permutation of $(0, 1, \dots, s-1)^T$, for $1\leq k \leq p$ and $1\leq j\leq \lambda$, and denote $\C=(\c_1,\dots,\c_p)$.
  \item[Step 4.]  Let $\widetilde{\D}_2=s\B+\C$, and obtain $\D_2=(\d_1,\dots,\d_{p})$ from
$\widetilde{\D}_2$ via the level-expansion method. Denote $\D=(\D_1,\D_2)$.
 \end{itemize}
 \end{construction}

%\begin{pro}\label{pro_dcd1}  The design $\D=(\D_1, \D_2)$   constructed by using $\D_1$, $\B$ and $\C$ in Construction \ref{con_dcd1} via (\ref{link_D1D2})  is a DCD$(\lambda s^2,s^q,p)$.
%\end{pro}

\begin{pro}\label{pro_dcd1}  The design $\D=(\D_1, \D_2)$   generated by Construction \ref{con_dcd1}  is a DCD$(\lambda s^2,s^q,p)$.
\end{pro}

The proof is straightforward, and thus we omit it.
In Construction \ref{con_dcd1}, Step 1 is devoted to constructing the $\D_1$ meets the requirement in Theorem \ref{iff1}. The orthogonal arrays $\A_1,\dots,\A_{\lambda}$ are
OA$(s^2,q+1,s,2)$'s. Note that  $\A_i$'s can be either the same or different (isomorphic or non-isomorphic), however, using different $\A_i$'s is more desirable for generating
$\D_1$ with
higher strength.
 The proposed procedure  produces
$(s!)^{\lambda s}\cdot(s!)^{\lambda}\cdot\lambda!$ different
quantitative columns in Proposition \ref{pro_dcd1}, that is,
Construction \ref{con_dcd1} provides the DCDs with a considerable
number of quantitative factors. {For $2\leq s\leq 11$ and a
positive integer $\lambda$, Construction 1 can produce the DCDs with
$\lambda s^2$ runs, $q$ qualitative factors and $p$ quantitative
factors, where $q=s$ for a prime power $s$, $q=2,~3$ for $s=6,~10$,
respectively, and $p\leq (s!)^{\lambda
s}\cdot(s!)^{\lambda}\cdot\lambda!$. Details are given in Table 1 of
the online supplementary material.}

Example \ref{ex2} illustrates Construction
\ref{con_dcd1}. To save the space, we set $p=3$.

 \begin{example} \label{ex2} Suppose we aim to construct a
 DCD$(27,3^3,3)$, that is,  $s=3, \lambda=3, q=3,
 p=3$. We use the
 three OA$(9,4,3,2)$'s below,
 \Bea
 \ba{ccc}
 \A_1=\left[\ba{cccc}
0&  0&  0&  0\\
1&  1&  2&  0\\
2&  2&  1&  0\\ \hdashline
0&  2&  2&  1\\
1&  0&  1&  1\\
2&  1&  0&  1\\ \hdashline
0&  1&  1&  2\\
1&  2&  0&  2\\
2&  0&  2&  2
 \ea\right],
  &
  \A_2=\left[\ba{cccc}
0&  0&  1&  0\\
1&  1&  0&  0\\
2&  2&  2&  0\\\hdashline
0&  2&  0&  1\\
1&  0&  2&  1\\
2&  1&  1&  1\\\hdashline
0&  1&  2&  2\\
1&  2&  1&  2\\
2&  0&  0&  2\\
 \ea\right],
 &
  \A_3=\left[\ba{cccc}
0&  0&  2&  0\\
1&  1&  1&  0\\
2&  2&  0&  0\\\hdashline
0&  2&  1&  1\\
1&  0&  0&  1\\
2&  1&  2&  1\\\hdashline
0&  1&  0&  2\\
1&  2&  2&  2\\
2&  0&  1&  2
 \ea\right].\ea
 \Eea
In Step 1, stack $\A_1,\A_2,\A_3$ by row and delete the
last column of the resulting design to obtain $\D_1$, which is an
OA$(27,3,3,2)$, and can be divided into 3 CROA$(9,3,3,2)$'s.
In Steps 2 and 3, let $\v_1=(1,2,0)^T, \v_2=(0,2,1)^T,
\v_3=(1,0,2)^T$, $\w_{11}=(0,1,2)^T$,
$\w_{12}=(1,0,2)^T$,
$\w_{13}=(0,2,1)^T$, $\w_{21}=(1,2,0)^T$,
$\w_{22}=(1,0,2)^T$, $\w_{23}=(0,1,2)^T$,
$\w_{31}=(2,0,1)^T$, $\w_{32}=(0,1,2)^T$
and $\w_{33}=(1,0,2)^T$,
then by Construction \ref{con_dcd1}, we can obtain
$$\B=\left(\setlength{\arraycolsep}{3pt} \ba{ccccccccccccccccccccccccccc}
1   &1  &1  &1  &1  &1  &1  &1  &1  &2  &2  &2  &2  &2  &2  &2  &2  &2  &0  &0  &0  &0  &0  &0  &0  &0  &0\\
0   &0  &0  &0  &0  &0  &0  &0  &0  &2  &2  &2  &2  &2  &2  &2  &2  &2  &1  &1  &1  &1  &1  &1  &1  &1  &1\\
1   &1  &1  &1  &1  &1  &1  &1  &1  &0  &0  &0  &0  &0  &0  &0  &0  &0  &2  &2  &2  &2  &2  &2  &2  &2  &2\ea\right)^T$$
and
$$\C=\left(\setlength{\arraycolsep}{3pt}
\ba{ccccccccccccccccccccccccccc}0   &0  &0  &1  &1  &1  &2  &2  &2  &1  &1  &1  &0  &0  &0  &2  &2  &2  &0  &0  &0  &2  &2  &2  &1  &1  &1\\
1   &1  &1  &2  &2  &2  &0  &0  &0  &1  &1  &1  &0  &0  &0  &2  &2  &2  &0  &0  &0  &1  &1  &1  &2  &2  &2\\
2   &2  &2  &0  &0  &0  &1  &1  &1  &0  &0  &0  &1  &1  &1  &2  &2  &2  &1  &1  &1  &0  &0  &0  &2  &2  &2\ea\right)^T.$$
In Step 4, let $\widetilde {\D}_2=3\B+\C$ and obtain $\D_2$.
According to Proposition \ref{pro_dcd1},
the resulting design $\D=(\D_1,\D_2)$ is a DCD$(27,3^3,3)$, listed in Table \ref{table_ex2}.
Moreover, the generated $\D_1$ is of strength 3.
The number of the qualitative factors in this
 example  achieves the upper bound in Corollary \ref{coro1}.

 \begin{table}[!h]\caption{$\D=(\D_1,\D_2)$ in Example \ref{ex2}}\label{table_ex2}
 \vspace{-6mm}
\bc {\small \tabcolsep=1.3pt \bt {r|rrrrrrrrrrrrrrrrrrrrrrrrrrr} \hline
\multirow{3}*{$\D_1^T$} &0 &1 &2 &  0&  1&  2&  0&  1&  2&  0&  1&  2&  0&  1&  2&  0&  1&  2&  0&  1&  2&  0&  1&  2&  0&  1&  2\\
~&0&    1&  2&  2&  0&  1&  1&  2&  0&  0&  1&  2&  2&  0&  1&  1&  2&  0&  0&  1&  2&  2&  0&  1&  1&  2&  0\\
~&0&    2&  1&  2&  1&  0&  1&  0&  2&  1&  0&  2&  0&  2&  1&  2&  1&  0&  2&  1&  0&  1&  0&  2&  0&  2&  1\\ \hline
\multirow{3}*{$\D_2^T$} & 9&    10&11&13&   14& 12& 15& 16& 17& 22& 23& 21& 19& 18& 20& 24& 25& 26& 2&  0&  1&  7&  8&  6&  4&  5&  3\\
                               ~& 3 &5& 4&  6&  7&  8&  0&  1&  2&  21& 22& 23& 19& 20& 18& 26& 24& 25& 11& 10& 9&  13& 14& 12& 16& 15& 17 \\
                                ~&16& 17&15&    10& 11& 9&  12& 13& 14& 1&  2&  0&  4&  5&  3&  8&  7&  6&  21& 22& 23& 19& 20& 18& 24& 25& 26\\\hline
 \et} \ec
\end{table}

 \end{example}

\begin{remark} If an OA$(\lambda s^2, q, s, t)$ with $t\geqslant 3$ can be partitioned into $\lambda$ CROA$(s^2, q, s, 2)$'s, a DCD with $\D_1$ of strength $t\geqslant 3$ can be constructed, such as the $\D_1$ in Example \ref{ex2}.
\end{remark}

We now introduce the second method to construct the required arrays $\D_1$, $\B$ and $\C$ in Theorem \ref{ns_con} based on the
permutation method.

\begin{construction} \label{con_dcd1_modi}
~
 \begin{itemize}
   \item[Step 1.] Obtain the array $\D_1$ by deleting the last column of $\one_{\lambda}\otimes \A_1$.
   \item[Step 2.]  Let $\B=(\b_1,\dots,\b_p)$, where $\{b_{i,k},b_{i+s^2,k},\dots,b_{i+(\lambda-1)s^2,k}\}$ is a random permutation of $\{0,1\dots,\lambda-1\}$
  and $b_{i,k}$ is the $i$-th entry of $\b_k$, for $1\leq i \leq s^2$ and $1\leq k\leq p$.

  \item[Step 3.]   Let ${\c}_k=\one_{\lambda}\otimes(\w_k\otimes \one_s)$, where ${\w}_{k}$ is a random permutation of $(0, 1, \dots, s-1)^T$, for $1\leq k\leq p$, and denote $\C=(\c_1,\dots,\c_p)$.

   \item[Step 4.]  Let $\widetilde{\D}_2 =s\B+\C$, and obtain $\D_2=(\d_1,\dots,\d_{p})$ from
$\widetilde{\D}_2$ via the level-expansion method. Denote $\D=(\D_1,\D_2)$.
 \end{itemize}
 \end{construction}

\begin{pro}\label{pro_case4}
The design $\D = (\D_1,\D_2)$ produced
 by  Construction \ref{con_dcd1_modi} is a  DCD$(\lambda s^2,s^q,p)$. \end{pro}

In Proposition \ref{pro_case4}, DCDs with at most $(s!)^{\lambda s}\cdot s!\cdot(\lambda!)^{s^2}$ distinct quantitative columns
can be generated, which indicates that Construction \ref{con_dcd1_modi}  can  also construct the DCDs containing a large number of quantitative factors.

Example \ref{ex_sa} below provides an illustration of Construction \ref{con_dcd1_modi}.
\begin{example}\label{ex_sa}
Consider generating a DCD$(27,3^3,3)$ and choose $\A_1$ shown in Example \ref{ex2}. In Step 1,
delete the last column of $\one_{\lambda}\otimes \A_1$ to obtain $\D_1$.
In Step 2, let
$$\B=\left(\setlength{\arraycolsep}{3pt}  \ba{ccccccccccccccccccccccccccc}2 & 0 & 1 &   2 & 0 & 1   & 2 & 0 & 1 & 0 & 1 & 2 & 0 & 1 & 2 &   0 & 1 & 2   & 1 & 2 & 0 &1  &2  &0  &1  &2  &0\\
2   &1  &0  &0  &2  &1  &1  &0  &2  &0  &2  &1  &1  &0  &2  &2  &1  &0  &1  &0  &2  &2  &1  &0  &0  &2  &1\\
0   &2  &1  &2  &1  &0  &1  &0  &2  &1  &0  &2  &0  &2  &1  &2  &1  &0  &2  &1  &0  &1  &0  &2  &0  &2  &1\ea\right)^T.$$
One can easily check that $\{b_{i,k},b_{i+9,k},b_{i+18,k}\}$ is a permutation of $\{0,1,2\}$, for $1\leq i \leq 9$ and $1\leq k\leq 3$. In Step 3, let $\w_1=(0,1,2)^T,\w_2=(1,2,0)^T$ and $\w_3=(2,0,1)^T$, and we have
$$\C=\left(\setlength{\arraycolsep}{3pt}  \ba{ccccccccccccccccccccccccccc}0 &0  &0  &1  &1  &1  &2  &2  &2  &0  &0  &0  &1  &1  &1  &2  &2  &2  &0  &0  &0  &1  &1  &1  &2  &2  &2\\
1   &1  &1  &2  &2  &2  &0  &0  &0  &1  &1  &1  &2  &2  &2  &0  &0  &0  &1  &1  &1  &2  &2  &2  &0  &0  &0\\
2   &2  &2  &0  &0  &0  &1  &1  &1  &2  &2  &2  &0  &0  &0  &1  &1  &1  &2  &2  &2  &0  &0  &0  &1  &1  &1\ea\right)^T.$$
The obtained design $\D=(\D_1,\D_2)$ is a DCD,  shown in Table \ref{table_ex_sa}.

 \begin{table}[!h]\caption{$\D=(\D_1,\D_2)$ in Example \ref{ex_sa}}\label{table_ex_sa}
\bc {\small \tabcolsep=1.3pt \bt {r|rrrrrrrrrrrrrrrrrrrrrrrrrrr} \hline
\multirow{3}*{$\D_1^T$}& 0  &1  &2  &0  &1  &2  &0  &1  &2  &0  &1  &2  &0  &1  &2  &0  &1  &2  &0  &1  &2  &0  &1  &2  &0  &1  &2\\
~&0 &1  &2  &2  &0  &1  &1  &2  &0  &0  &1  &2  &2  &0  &1  &1  &2  &0  &0  &1  &2  &2  &0  &1  &1  &2  &0\\
~&0&    2   &1  &2  &1  &0  &1  &0  &2  &0  &2  &1  &2  &1  &0  &1  &0  &2  &0  &2  &1  &2  &1  &0  &1  &0  &2\\ \hline
\multirow{3}*{$\D_2^T$} &19 &1  &9  &22 &3  &13 &25 &8  &17 &0  &11 &18 &5  &14 &21 &6  &16 &24 &10 &20 &2  &12 &23 &4  &15 &26 &7\\
~&23    &12 &4  &7  &26 &17 &10 &1  &19 &3  &22 &14 &16 &8  &25 &18 &9  &0  &13 &5  &21 &24 &15 &6  &2  &20 &11\\
~&8 &26 &17 &20 &10 &2  &13 &5  &21 &16 &7  &25 &1  &19 &11 &22 &14 &3  &24 &15 &6  &9  &0  &18 &4  &23 &12\\\hline
 \et} \ec
\end{table}

\end{example}

In practice, for a
 predetermined $p$, an optimal $\D_2$
according to some optimization criteria
(such as maximin distance, uniform discrepancies, etc)
can be found by ranking all possible candidate designs or via the greedy search algorithms such as the simulated annealing or the threshold accepting algorithms if the number of candidate designs is exceedingly large (\cite{MM95,BMB15}).

%{The optimal algorithm of Construction  \ref{con_dcd1} and the optimal designs are presented in the online supplementary material.}

%\subsection{Constructions for better space-filling property on the quantitative factors}\label{sub_con2}

\vspace{2mm}\noindent {\bf 4.2 Constructions for the better space-filling property on the quantitative factors}\vspace{2mm}

This subsection provides another construction method which uses one array we call $\A$ to provide $\D_1$ and $\C$ that is required in Theorem \ref{ns_con}. That is, the new construction only involves two arrays $\A$ and the $\B$.  Two specific
cases of the construction are provided to produce the required $\A$ and
$\B$, where the resulting DCDs
may share some extra high-dimensional
space-filling properties among the quantitative factors. Suppose ${\A} $ is an  OA$(n,q+1,s,2)$
and $\B $ is an OA$(n,p,{n}/{s^2},1)$. Construction 3 works as follows.

%When the pairs of columns of $\A$ and any column of $\B$ in Construction \ref{con_general} meet  the following condition, the DCDs can be constructed.

\begin{construction}\label{con_general}
~
 \begin{itemize}
  \item[Step 1.] Randomly choose $q$ columns from $\A$ to obtain
 $\D_1$. Denote the remaining column of $\A$ by ${\bf a}^*$.
\item[Step 2.]   Let $\C=(\c_1,\dots,\c_p)$,  where  $\c_k$ is obtained by permuting the levels of ${\bf a}^*$, for any $1\leq k\leq p$.
\item[Step 3.]   Let $\widetilde{\D}_2=s\B+\C$, and obtain $\D_2=(\d_1,\dots,\d_{p})$ from
$\widetilde{\D}_2$ via the level-expansion method. Denote $\D=(\D_1,\D_2)$.
\end{itemize}
\end{construction}

\begin{theorem}\label{pro_general}
Suppose ${\A}=({\bf a}_1,\dots,{\bf a}_{q+1}) $ is an  OA$(n,q+1,s,2)$
and $\B=(\b_1,\dots,\b_p)$ is an OA$(n,p,{n}/{s^2},1)$. If $
({\bf a}_i,{\bf a}_j,\b_k)$ is an OA$(n,3,s^2\left({n}/{s^2}\right),3)$ for any $1 \leq i\neq j \leq q+1$ and $ 1 \leq k\leq p$, then the obtained $\D$ by Construction \ref{con_general} above is a DCD$(n,s^{q},p)$. \end{theorem}

Theorem \ref{pro_general} tells us that, to
construct a DCD$(n,s^q,p)$, the most important task is to find such two required arrays $\A$ and $\B$. Under the condition of Theorem \ref{pro_general}, it can be verified that the three arrays $\D_1$, $\B$ and $\C$ in Construction \ref{con_general} meet the conditions in Theorem \ref{ns_con}.
Hence, the obtained  $\D$  of Construction \ref{con_general} is a DCD. One can see that Theorem \ref{pro_general} can be regarded as a special case of Theorem \ref{ns_con}. %{The potential benefit of Theorem \ref{pro_general} is that we only need to find two certain arrays $\A$ and $\B$ instead of three arrays $\D_1$, $\B$ and $\C$ like Theorem \ref{ns_con}, since the two arrays $\D_1$ and $\C$ can be generated from the same orthogonal array, $\A$, according to Steps 1 and 2 of Construction \ref{con_general}}.

Next, we present an example to illustrate the application of Construction \ref{con_general}.
\begin{example}\label{ex3} Suppose that
we want to construct a DCD($8,2^2,4$).
%an 8-run DCD with two qualitative two-level factors and four quantitative
%factors, i.e., $\D_1$ {is an} OA$(8,2,2,2)$ and $\D_2
%$ {is an LH}$(8,4)$.
Let $$\A =({\bf a}_1,{\bf a}_2,{\bf a}_3)= \left( \ba{ccc}
0&  0&  0\\
0&  1&  1\\
1&  0&  1\\
1&  1&  0\\
0&  0&  0\\
0&  1&  1\\
1&  0&  1\\
1&  1&  0
 \ea \right) \text{and}~
\B =(\b_1,\b_2,\b_3,\b_4)= \left( \ba{cccc}
0&  0&  0&  0\\
0&  1&  1&  0\\
1&  0&  1&  0\\
1&  1&  0&  0\\
1&  1&  1&  1\\
1&  0&  0&  1\\
0&  1&  0&  1\\
0&  0&  1&  1
 \ea\right).$$
It can be checked that \A~is an OA$(8,3,2,2)$, $\B$ is an OA$(8,4,2,2)$, and
$({\bf a}_i,{\bf a}_j,\b_k)$ is an OA$(8,3,2,3)$, for
$1\leq i\neq j\leq 3,1\leq k\leq 4$. Thus,  \A~and $\B$  satisfy the requirements in Theorem \ref{pro_general}.
In Step 1, select ${\bf a}_2,{\bf a}_3$ of \A~to be $\D_1$ and denote ${\bf a}^*={\bf a}_1$.
In Step 2, each column of $\C$ is generated by ${\bf a}_1$ via the level permutation. Without loss of generality, let $\c_k={\bf a}_1$ for $1 \leq k \leq 4$.
In Step 3,
we can obtain the corresponding  $\widetilde{\D}_2 $ and apply the level-expansion method. The resulting design $\D=(\D_1,\D_2)$ is shown in Example \ref{ex1}.
Additionally, $\B$ is an orthogonal array of strength
2, therefore, the resulting $\D_2$ achieves the
 stratifications on $2\times 2$ grids for any two quantitative factors, which can be
verified by Figure \ref{figure_ex1}.

\end{example}

We now present two cases to generate the required arrays $\A$ and $\B$ in Theorem \ref{pro_general},  and they can produce DCDs with $s^3$ and $s^u $ runs for $u\geqslant 3$, respectively. Besides, Case \ref{con1}
is suitable for any $s\geqslant 2$, while Case
 \ref{case_regular_general} works for
any prime power $s$.
In both cases,
the resulting DCDs enjoy some extra two-dimensional
 space-filling properties among the quantitative factors.
\begin{case}\label{con1}
Let $\G$ be an $OA(s^3,m,s,3)$. Split the columns of $\G$ randomly into two arrays, $\A$ and $\B$, where $\A$ has $q+1$ columns and $\B$ has $p$ columns, $m = p+q+1$.
\end{case}

The orthogonal array $\G$ of strength 3 in Case
\ref{con1} can be directly
taken from the existing websites, such as \cite{S14}.

\begin{pro}\label{pro_con_case1}
 The $\D=(\D_1,\D_2)$ constructed via Construction \ref{con_general} by using $\A$ and $\B$ in Case  \ref{con1}   is a DCD$(s^3,s^q,p)$, where $q+p=m-1$. Furthermore, we have
 \begin{enumerate}
\item[(a)] $\D_1$ is an OA$(s^3,q,s,t)$, where $t=q$, if $q<3$,  and $t=3$, if $q\geqslant 3$;

\item[(b)] $(\tilde{\d_k},\tilde{\d_{k'}}),$ for any $1\leq k \neq
k' \leq p$, achieves the stratification on $s^2\times s$ and $s \times s^2$ grids; and

\item[(c)] $\widetilde{\widetilde{\D}}_2$ is an OA$(s^3,p,s,t)$, where $t=p$, if $p<3$,  and $t=3$, if $p\geqslant 3$.
\end{enumerate}
\end{pro}

Parts (b) and (c) in Proposition \ref{pro_con_case1} mean
that $\D_2$ enjoys the two-dimensional and
three-dimensional space-filling properties. {For $s\leq 10$, the sum of the number of the qualitative and quantitative factors of the DCDs produced by Case 1 of Construction 3 is no more than three for $s=2,3,6,10$, five for $s=4,5$, seven for $s=7$, and nine for $s=8, 9$. Table 2 of the online supplementary material shows the details.}

We now begin to introduce Case \ref{case_regular_general}, which is based on regular fractional factorial designs (\cite{WH09}).
For any prime power $s$ and any integer $u \ge 3$, let
 $\Xi_1,\dots,\Xi_u$
be independent columns of length $s^u$ with the entries being from $GF(s)$, the Galois field of order $s$.

\begin{case}\label{case_regular_general}
~
\begin{itemize}
\item[Step 1.] Let
\Bea%\label{groups}
 \begin{split}
  &\A=\left\{\Xi_1+\mu_2\Xi_2 ~|~ \mu_2\in GF(s) \right\}\cup\left\{\Xi_2\right\}=({\bf a}_1,\dots,
{\bf a}_{s+1}),\\
 &\R_{v}=\{\Xi_1+\mu_2\Xi_2+\mu_{v+2}\Xi_{v+2}~|~\mu_2\in GF(s), \mu_{v+2} \in GF(s)\backslash \{0\}\}\\
&~~~~~~~~ \cup\{\Xi_2+\mu_{v+2}\Xi_{v+2}~|~\mu_{v+2} \in GF(s)\backslash \{0\}\}\\
&~~~~~~~~ \cup\{\Xi_{v+2}\} =(\r_{v,1},\dots, \r_{v,s^2}),
\end{split}
\Eea
where $\r_{v,f}$ is a column vector of length $s^u$, for
$1\leq v\leq u-2$ and $1\leq  f\leq s^2$.

\item[Step 2.] For any $1\leq f\leq s^2,$ let
\Bea \label{F_f} \B_f=(\r_{1,f},\dots,\r_{u-2,f})\T,
 \Eea
 where
  $$\T=\left(\ba{ccccc} s^{u-3} & 1& \cdots & s^{u-5} & s^{u-4} \\
s^{u-4}  & s^{u-3} & \cdots & s^{u-6} & s^{u-5} \\
\vdots &  \vdots & \vdots & \vdots & \vdots \\
s & s^2 & \cdots & s^{u-3} & 1\\
1 & s & \cdots & s^{u-4} & s^{u-3}
\ea\right)=(\t_1,\dots,\t_{u-2}).$$
There are $u-2$ columns in each $\B_f.$

\item[Step 3.] Let $\B=(\B_1,\dots,\B_{s^2})=(\b_1,\dots,\b_{(u-2)s^2})$.
\end{itemize}
\end{case}

Clearly, $\A$ consists of the independent
columns $\Xi_1,\Xi_2$ and all possible interactions of these
 two columns, and thus $\A$ has $s+1$ columns.
While the column vectors in $\R_v$ must involve
$\Xi_{v+2}$ and may contain $\Xi_1$ and $\Xi_2$. Lemma \ref{le1} summarizes the design properties of  $\A,\R_1,\dots,\R_{u-2}$.
The proof is straightforward and thus omitted.

\begin{lemma}\label{le1} For   $\A,\R_1,\dots,\R_{u-2}$  in Case \ref{case_regular_general}, we have,
\begin{enumerate}
\item[(a)] $\A$ is an OA$(s^u,s+1,s,2)$;
\item[(b)] $(\R_1, \dots, \R_{u-2})$ is an OA$(s^u,(u-2)s^2,s,2)$;
\item[(c)] $({\bf a}_i,{\bf a}_j,\r_{1,f},\dots,\r_{u-2,f})$ is an OA$(s^u,u,s,u)$, for any $ 1\leq f\leq s^2$, $1 \leq i\neq j \leq s+1$; and
\item[(d)] $(\r_{1,f},\dots,\r_{u-2,f},\r_{v,l})$ is an OA$(s^u, u-1, s, u-1)$,  for any
$1\leq v\leq u-2$, $1 \leq f\neq l\leq s^2$.
\end{enumerate}
\end{lemma}

Lemma \ref{le1}(c) means that taking two distinct
columns from $\A$, and one column from each
$\R_v$, for $v =1, \ldots, u-2$, the resulting $u$
columns form an $s$-level orthogonal array of $s^u$
runs and strength $u$, that is,  a full factorial design of
$s$ levels and $u$ columns.
Similarly, Lemma \ref{le1}(d) implies that
the array of $u-1$ columns, consisting of two distinct
columns of $\R_v$ and one column of each of the remaining $u-3$ arrays $\R_1,\dots, \R_{v-1}, \R_{v+1},\dots,\R_{u-2}$, is an orthogonal array of strength $u-1$, i.e.,
a full factorial design of $s$ levels and $u-1$ columns.

From Lemma \ref{le1} and  Case
\ref{case_regular_general}, we have the following result.

\begin{lemma} \label{regular_le3}
For $\A$  and $\B$ in Case \ref{case_regular_general}, we have
\begin{itemize}
\item[(a)] $({\bf a}_i,{\bf a}_j,\b_k)$ is an OA$(s^u,3,s^2(s^{u-2}),3)$,
 for any $1\leq i\neq j \leq s+1$ and $1\leq k\leq (u-2)s^2$; and
\item[(b)] $\A $ is an $ {\rm OA}(s^u,s+1,s,2)$ and $\B$ is an {\rm OA}$(s^u,(u-2)s^2,s^{u-2},1)$.
\end{itemize}
\end{lemma}

Lemma \ref{regular_le3} points out that $\A$ and $\B$ in Case \ref{case_regular_general}  are the required arrays
in Theorem \ref{pro_general}. For $s=2, u=3$, the two arrays $\A$
 and $\B$ in Case \ref{case_regular_general}
are shown in Example \ref{ex3}.

\begin{pro}\label{pro_case_regular_general}
For any prime power $s$ and any integer $u\geqslant3$,  $\D = (\D_1,\D_2)$ constructed via Construction \ref{con_general} by
using $\A$ and $\B$ in Case
\ref{case_regular_general} is  a DCD$(s^u,s^s,(u-2)s^2)$
with $\D_1$ being an  OA$(s^u,s,s,2)$
and $\D_2$ being an LH$(s^u,(u-2)s^2)$.  In addition, $\D_2$ has the following properties
\begin{enumerate}
\item[(a)] if $\lfloor (i-1)/(u-2) \rfloor=\lfloor (i'-1)/(u-2) \rfloor$,  $\tilde{\tilde{\d_i}}$ and $\tilde{\tilde{\d_{i'}}}$ achieve $s\times s$ grids stratification; and
\item[(b)]   if $\lfloor (i-1)/(u-2)\rfloor \neq \lfloor (i'-1)/(u-2)\rfloor$,
 $\tilde{\tilde{\d_i}}$ and $\tilde{\tilde{\d_{i'}}}$  achieve $s^{u-2}\times s$  and $s\times s^{u-2}$ grids stratification.
\end{enumerate}
\end{pro}

{Obviously, the number of the qualitative factors for the DCDs in Proposition
 \ref{pro_case_regular_general} is $s$, which reaches the upper bound in
Corollary \ref{coro1} and the number of the quantitative factors is $(u-2)s^2$. }

\section{Conclusion}\label{conc}

In this paper, we propose $\omega$-way coupled designs with $\omega\geqslant 2$, for computer experiments involving both qualitative and quantitative
factors. We focus on the properties and constructions of the two-way coupled designs, namely, doubly coupled designs. Similar to MCDs, such designs are an economical alternative to SLHDs. Different from MCDs, they require that for each level combination of every two qualitative factors, the corresponding design points for quantitative factors form an LHD.  This additional requirement leads to the result that given the same run size, DCDs can entertain less qualitative factors than MCDs.
 In addition, DCDs are equipped with the better stratification
properties between the qualitative and quantitative factors than MCDs.

When the design for the
qualitative factors $\D_1$ is an OA$(n,q,s,2)$, the necessary and
sufficient
conditions for the existence of a DCD
are provided and a tight upper bound of the number of
the qualitative factors is given. Three construction methods are provided.  They are different but related.  Particularly, Constructions \ref{con_dcd1}  and \ref{con_dcd1_modi}  are both based on the idea of permutations, but they generate $\D_1, \B, \C$ in different ways.  More specifically,   Step 1 of Construction \ref{con_dcd1_modi} uses $\lambda$ identical OAs while $\A_1,\cdots,  \A_{\lambda}$ used in Step 1 of Construction \ref{con_dcd1}   can be identical, isomorphic or non-isomorphic; Step 2 of Construction  \ref{con_dcd1}  is a special case of Step 2 of Construction \ref{con_dcd1_modi} when $\{b_{i,k},b_{i+s^2,k},\dots,b_{i+(\lambda-1)s^2,k}\}$  is the same random permutation of $\{0,1,\cdots,\lambda-1\}$ for $1 \leq i \leq s^2$; Step 3 of Construction 2 is a special case of Step 3 of Construction 1 when $\omega_{kj}$ is the same permutation of $\{0, 1, \cdots, \lambda -1\}$ for $1 \leq j \leq \lambda$.  Construction 3 is different from Constructions \ref{con_dcd1}  and \ref{con_dcd1_modi} in that it uses an array $\A$ to provide $\D_1$ and $\C$.  Thus the building block of Construction 3 is arrays $\A$ and $\B$ that meet the conditions in Theorem~\ref{pro_general}.  Two cases of such $\A$'s and $\B$'s are given. Because $\B$'s in Case 1 and $\R$'s in Case 2 are orthogonal arrays,  the
 $\D_2$'s of DCDs produced by Construction 3 involving Cases 1 and 2 are orthogonal array-based Latin hypercubes.   Constructions \ref{con_dcd1} and \ref{con_dcd1_modi} can
 entertain a large number of the quantitative
 factors than Construction \ref{con_general}, but their limitation   is that the space-filling properties of the designs for the quantitative factors cannot be ensured. On the other hand,  the resulting
 $\D_2$ constructed by Construction \ref{con_general} along with Case \ref{con1} and Case \ref{case_regular_general} can guarantee some desirable stratification on grids for
 the quantitative factors, however, the number of the quantitative factors of the resulting designs may be relatively limited. As all the constructions are algebraic, they do not cost computing time.  For the practical use, we list examples of DCDs provided by the proposed construction methods in the online supplementary material.

The needed arrays $\D_1$, $\B$ and $\C$ in Theorem \ref{ns_con} can be constructed by other approaches in the future. The methods to generate the two arrays $\A$ and $\B$ required by Theorem \ref{pro_general}  are not limited to the two
cases given in this paper, and more pairs of
$\A$ and $\B$ can be considered  in the future. Furthermore, we can consider the
DCD, $\D=(\D_1,\D_2)$ with $\D_1$ being a mixed-level orthogonal array, being an OA of strength $t$, or possessing some good space-filling properties.  Study of the space-filling property of $\D_1$ is an important topic. The work in \cite{ZX14} studied the space-filling property of orthogonal arrays under two commonly used space-filling measures, discrepancy and maximin distance. Because of the requirement relationship between columns in $\D_1$ and columns in $\D_2$,  it would require additional effort to explore the theoretical space-filling property of $\D_1$ in a DCD.
Another possible direction is to construct DCDs, where $\D_2$ has the high-dimensional space-filling properties, such as 3 to 4 dimensions.  In addition, an interesting but challenging direction is to construct $\omega$-way coupled designs with $\omega>2$.  The construction of such designs is not trivial and cannot be easily extended. We hope to study this and report the results in the near future.

\vskip 14pt
\noindent {\large\bf Acknowledgements}

Partial of the research was done when the first author visited the
Department of Mathematics and Statistics at Queen's University. Yang
is supported by the China Scholarship Council, Research start-up
funding of Sichuan Normal University  and National Natural Science
Foundation of China (12101435). Zhou is supported by the National
Natural Science Foundation of China (11871288) and Natural Science
Foundation of Tianjin (19JCZDJC31100). Lin is supported by the
Natural Sciences and Engineering Research Council of Canada. The
authors would also like to thank Professor Jianfeng Yang for his
valuable suggestions on this topic.

\bibhang=1.7pc
\bibsep=2pt
\fontsize{9}{14pt plus.8pt minus .6pt}\selectfont
\renewcommand\bibname{\large \bf References}

%\begin{thebibliography}{11}
\expandafter\ifx\csname
natexlab\endcsname\relax\def\natexlab#1{#1}\fi
\expandafter\ifx\csname url\endcsname\relax
  \def\url#1{\texttt{#1}}\fi
\expandafter\ifx\csname urlprefix\endcsname\relax\def\urlprefix{URL}\fi
%\fi

\lhead[\footnotesize\thepage\fancyplain{}\leftmark]{}\rhead[]{\fancyplain{}\rightmark\footnotesize\thepage}%Put this line in Page 2
%\markboth{\hfill{\footnotesize\rm Feng Yang and C. Devon Lin}
%\hfill} {\hfill {\footnotesize\rm DOUBLY COUPLED DESIGNS FOR COMPUTERS EXPERIMENTS} \hfill}
%%%%%%%%%%%%%%%%%%%%%%%%%%%%%%%%%%%%%%%%%%%%%%%%%%%%%%%%%%

\vskip .65cm
\noindent
School of Mathematical Sciences \& Laurent Mathematics Center, Sichuan Normal University, Chengdu 610066, China
\vskip 2pt \noindent E-mail:  yangfeng@sicnu.edu.cn
\vskip 2pt

\noindent Department of Mathematics and Statistics, Queen's University, Kingston, ON, K7L 3N6, Canada \vskip 2pt \noindent E-mail: devon.lin@queensu.ca

\noindent School of Statistics and Data Science, LPMC \& KLMDASR, Nankai University, Tianjin
300071, China \vskip 2pt \noindent E-mail: ydzhou@nankai.edu.cn

\noindent School of Statistics, Beijing Normal University, Beijing 100875, China \vskip 2pt

\noindent E-mail: heyuanzhen@bnu.edu.cn
% \vskip .3cm
%\centerline{(Received ???? 20??; accepted ???? 20??)}\par

\end{document}